\documentclass[10pt]{article}
\usepackage{bbm}
\usepackage[final]{graphics}
\usepackage{amsmath}
\usepackage{amsfonts,amsbsy}
\usepackage{amssymb}
\usepackage[dvips]{color}
\def\empile#1\over#2{\mathrel{\mathop{\kern 0pt#1}\limits_{#2}}}

\def\bs{\boldsymbol}

\newcommand{\slv}{\raise.15ex\hbox{$/$}\kern-.53em\hbox{$v$}}
\newcommand{\slF}{\raise.15ex\hbox{$/$}\kern-.53em\hbox{$F$}}
\newcommand{\slL}{\raise.15ex\hbox{$/$}\kern-.53em\hbox{$L$}}
\newcommand{\slP}{\raise.15ex\hbox{$/$}\kern-.53em\hbox{$P$}}
\newcommand{\slp}{\raise.15ex\hbox{$/$}\kern-.53em\hbox{$p$}}
\newcommand{\slq}{\raise.15ex\hbox{$/$}\kern-.53em\hbox{$q$}}
\newcommand{\slR}{\raise.15ex\hbox{$/$}\kern-.53em\hbox{$R$}}
\newcommand{\slQ}{\raise.15ex\hbox{$/$}\kern-.53em\hbox{$Q$}}
\newcommand{\slK}{\raise.15ex\hbox{$/$}\kern-.53em\hbox{$K$}}
\newcommand{\slk}{\raise.15ex\hbox{$/$}\kern-.53em\hbox{$k$}}
\newcommand{\slD}{\raise.15ex\hbox{$/$}\kern-.53em\hbox{$D$}}
\newcommand{\slC}{\raise.15ex\hbox{$/$}\kern-.53em\hbox{$C$}}
\newcommand{\slA}{\raise.15ex\hbox{$/$}\kern-.53em\hbox{$A$}}
\newcommand{\slSigma}{\raise.15ex\hbox{$/$}\kern-.53em\hbox{$\Sigma$}}
\newcommand{\slpartial}{\raise.15ex\hbox{$/$}\kern-.53em\hbox{$\partial$}}
\newcommand{\slcalP}{\raise.15ex\hbox{$/$}\kern-.63em\hbox{$\cal P$}}

\def\q{{\boldsymbol q}}
\def\l{{\boldsymbol l}}
\def\k{{\boldsymbol k}}

\def\x{{\boldsymbol x}}
\def\y{{\boldsymbol y}}

\def\r{{\boldsymbol r}}
\def\z{{\boldsymbol z}}
\def\v{{\boldsymbol v}}
\def\w{{\boldsymbol w}}

\def\u{{\boldsymbol u}}

\catcode`\@=11

\newcount\@tempcntc
\def\@citex[#1]#2{\if@filesw\immediate\write\@auxout{\string\citation{#2}}\fi
  \@tempcnta\z@\@tempcntb\m@ne\def\@citea{}\@cite{%
        \@for\@citeb:=#2\do%
    {\@ifundefined{b@\@citeb}%
        {\@citeo\@tempcntb\m@ne\@citea%
                \def\@citea{,\penalty\@m\ }{\bf ?}\@warning%
                {Citation `\@citeb' on page \thepage \space undefined}}%
        {\setbox\z@\hbox{\global\@tempcntc0\csname b@\@citeb\endcsname\relax}
     \ifnum\@tempcntc=\z@ \@citeo\@tempcntb\m@ne%
       \@citea\def\@citea{,\penalty\@m}%
       \hbox{\csname b@\@citeb\endcsname}%
     \else%
      \advance\@tempcntb\@ne%
      \ifnum\@tempcntb=\@tempcntc%
      \else\advance\@tempcntb\m@ne\@citeo%
      \@tempcnta\@tempcntc\@tempcntb\@tempcntc\fi\fi}}\@citeo}{#1}}%

\def\@citeo{\ifnum\@tempcnta>\@tempcntb\else\@citea
  \def\@citea{,\penalty\@m}%
  \ifnum\@tempcnta=\@tempcntb\the\@tempcnta\else
   {\advance\@tempcnta\@ne\ifnum\@tempcnta=\@tempcntb \else
\def\@citea{--}\fi
    \advance\@tempcnta\m@ne\the\@tempcnta\@citea\the\@tempcntb}\fi\fi}

\catcode`\@=12

\begin{document}

\title{\bf Relating the description of gluon production in
    \textit{p}\textit{A} collisions and parton energy loss in
    \textit{A}\textit{A} collisions}
\author{Yacine Mehtar-Tani$^{(1)}$}
\maketitle
\begin{center}
\begin{enumerate}

\item Laboratoire de Physique Th\'eorique\\
  Universit\'e Paris Sud, B\^at. 210\footnote{Unit\'e mixte de recherche du CNRS (UMR 8627)}\\
  91405, Orsay cedex
\end{enumerate}
\end{center}

\begin{abstract}
 We calculate the classical gluon field of a fast projectile passing through a dense medium. We show that this allows us to calculate both the initial state gluon production in proton-nucleus collisions and the final state gluon radiation off a hard parton produced in nucleus-nucleus collisions. This unified description of these two phenomena makes the relation between the saturation scale $Q_s$ and the transport coefficient $\hat q$ more transparent. Also, we discuss the validity of the eikonal approximation for gluon propagation inside the nucleus in proton-nucleus collisions at RHIC energy.

\end{abstract}
\vskip 5mm
\begin{flushright}
LPT-Orsay 06-41
\end{flushright}
\section{Introduction}
The advent of a new generation of colliders, RHIC and LHC, has stimulated the development of new tools for the understanding of high energy and high density systems. The purpose of studying heavy ion collisions at high energy is to create such dense systems, leading eventually to evidences for the existence of the Quark-Gluon-Plasma (QGP) predicted by the QCD phase diagram. Very early, it appeared that one has to distinguish between final state interactions, occurring after the hard parton production, and initial state interactions responsible for hard parton production in nucleus-nucleus collisions; this motivated the study of deuteron-gold collisions at RHIC where final state interaction are absent. \\
The physics of proton-nucleus collisions at high energy turned out to be very rich in new features compared to proton-proton collisions, for instance: high $p_t$ suppression at forward rapidities, Cronin enhancement (also observed at low energy), centrality dependence of spectra, etc. \cite{BRAHMS1,exp}. These results have been confronted with the theory of the Color Glass Condensate (CGC) which describes the nuclear wave function at high energy \cite{CGCrev1,CGCrev2}. Basically, it extends small coupling QCD calculations to a region, the so-called saturation regime, characterized by the hard scale $Q_s$ (saturation scale), where high density effects do not allow one to apply the usual perturbative QCD. \\
    ~~~~Historically the idea of saturation of the gluon distribution at high energy has been introduced in the early 80's \cite{GriboLR1,MuellQ1,BlaizM1} as a necessary condition for unitarity. Indeed at high energy gluons dominate the dynamics and the growth of the number of gluons, driven by the BFKL evolution equation, violates unitarity. This equation resums large $\log(s)$ effects but disregards gluon recombination \cite{BalitL1,KuraeLF1}. A first version of the CGC known as the McLerran-Venugopalan model pointed out the interest of the semi-classical picture for the description of high density systems. \cite{McLerV1,McLerV2,McLerV3}. A couple of years later, a more sophisticated theory has been built which incorporates the main BFKL features and extends them to the non-linear regime: the saturation regime, leading to the BK-JIMWLK equation \cite{IancuLM1,IancuLM2,FerreILM1,JalilKLW1,JalilKLW2,JalilKLW3,JalilKLW4,KovneM1,KovneMW3,JalilKMW1,Balit1,Kovch3}.\\
    ~~~~On the other hand, several works inspired by the Landau-Pomeranchuk-Migdal (LPM) effect in QED \cite{LPM1,LPM2,LPM3}: the BDMPS formalism, based on Feynman diagrams calculations \cite{BDMPS1,BDMPS2,BDMPS3,BDMPS4,BDMPS5,BDMPS6,BDMPS7},  and the Zakharov formalism (Z) based on path integrals \cite{Zak1,Zak2,Zak3,Zak4,Zak5,Zak6}, have been dedicated to the study of final state interactions in nucleus-nucleus collisions, especially to understand the large suppression of observed large $p_t$ hadron spectra compared to proton-proton collisions. The basic idea is to explain this suppression by the  energy loss of the initially produced hard parton. It looses part of its energy by radiating gluons when passing through the dense medium formed after the collision. In \cite{Wied1,Wied2,Wied3,Wied4}, Wiedemann (W) has provided another treatment of the problem also in terms of Feynman diagrams.\\
    In the present work, we give a new and simple derivation of the radiated gluon spectrum, recovering the BDMPS-Z-W result. We provide a universal formulation for both gluon production in proton-nucleus collisions \cite{GM}, and gluon radiation off a produced hard parton in nucleus-nucleus collisions. The basic idea is to calculate gluon radiation off a high energy projectile passing through a dense medium in the semi-classical picture. To do that, we solve the Yang-Mills equations for the radiated gluon field $\delta A^\mu$, which is treated as a perturbation of the background medium field $A^\mu_0$. This medium could be a nucleus (cold matter) in proton-nucleus collisions or a hot medium produced after the collision of two heavy nuclei. This approach turns out to be a very useful framework, which avoids many  technical problems, making the picture clear. Obviously, the obtained classical field is a function of the medium source density $\rho_0$ (or equivalently, the medium background field $A^\mu_0$). The source distribution $\rho_0$ is a random quantity. Thus, to calculate observables one has to average over all possible source configurations with a given statistical weight ${\cal W}[\rho_0]$:
    \begin{equation}
    \langle{\cal O}\rangle\equiv\int {\cal D}[\rho_0] {\cal W}[\rho_0]{\cal O}[\rho_0].
    \end{equation}
       \\
       The differential average number of produced gluons is given by the formula:
       
       \begin{equation}
       \omega\frac{d\langle N \rangle}{d\omega d^2\q_\perp}=\frac{1}{16\pi^3} \langle \sum_\lambda|\int d^4x \square_x \delta A_\mu(x) \epsilon^\mu_{(\lambda)}(\q) e^{ix.q}|^2   \rangle   \label{eq:dN}
       \end{equation}
       where $\epsilon^\mu_{(\lambda)}$ is the polarization vector of the radiated gluon, and $q\equiv(q^+=\omega,q^-=\q_\perp^2/2\omega,\q_\perp)$ its momentum.\\
    ~~~~In section 2, we derive the gauge field induced after the interaction of a fast projectile with an unspecified medium characterized by the statistical weight $\cal W$, by solving the Yang-Mills equations. Choosing the gauge as the light cone gauge of the projectile allows us to give a simple derivation, and we show, in section 3, that it is straightforward to deduce from the classical field the induced radiative gluon spectrum in nucleus-nucleus collisions, and in the approximation of independent scattering centers we reproduce a well-known formula derived in the more formal approach mentioned above (see BDMPS-Z-W papers listed above). \\
    In section 4, we consider gluon production in proton-nucleus collisions. Assuming the interaction time to be much smaller than any other time scale appearing in the problem (that amounts to consider gluon propagation inside the nucleus to be eikonal) we recover the $k_t$-factorization formula which resums high density effects in the nucleus \cite{GM,BGV1,KovnW,KovchM3,DumitM1,KovchT1}. Then we discuss the validity of this approximation, and show that it fails in some kinematic regime probed at RHIC energies. 
    Finally, we summarize in section 5.   

\section{The gauge field of a fast moving parton (or hadron) passing through  a dense medium}
\begin{figure}[ht]
\begin{center}
\resizebox*{!}{4cm}{\includegraphics{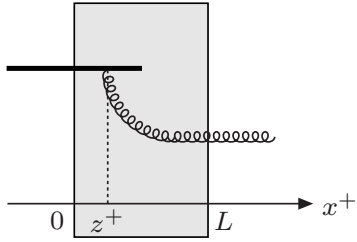}}
\caption{A schematic representation of eq. (\ref{eq:field}). The fast projectile (thick line) passes through the medium of thickness $L$ and emits a gluon at the time $z^+$. The gluon emission could also occur outside the medium, at $z^+<0$ or $z^+>L$.  }\label{fig1}
\end{center} 
\end{figure}

We consider a massless parton (or a hadron described as a collection of partons) moving with the velocity of light, in the $x^+$ direction. At some time it passes through a dense static medium of size $L$ \footnote{The assumption of a static medium (static scattering centers) in parton energy loss has been used in BDMPS-W-Z formalism and in the Gyulassy-Wang model \cite{GW}.}. The medium is described by the following current, in the medium rest frame, 
\begin{equation}
J_{0(r.f)}^\mu=\rho_0(x^3,\x_\perp) \delta^{\mu 0}.
\end{equation}
We are interested in the induced radiative spectrum off the hard parton inside the medium. Knowing the simplicity of solving the Yang-Mills equations in the light-cone gauge of the parton $A^+=0$ \cite{GM}, we perform a boost of velocity $\beta\sim-1$ (the cross-section is Lorentz invariant) which affects only the medium, namely the parton remains in the $x^+$ direction and the medium is pushed very close to the light-cone in the $x^-$ direction. \\    
In the boosted frame the medium field can be checked to be \footnote{For any gauge choice for the medium field in its rest frame the $A^+$ component is suppressed by the boost.} \cite{BGV1}:
  \begin{equation}
A^\mu_0=-\delta^{-\mu}
\frac{1}{{\bs\partial}_\perp^2}\rho_0(\x_\perp,x^+)\;\label{eq:A-} ,
\end{equation} 
In the light-cone gauge of the fast parton, $A^+=0$, the Yang-Mills equations read
\begin{eqnarray}
&&
-\partial^+(\partial_\mu A^\mu)-ig[A^i,\partial^+A^i]=J^+,
\nonumber\\
&&
[D^-,\partial^+A^-]-[D^i,F^{i-}]=J^-,
\nonumber\\
&&
\partial^+F^{-i}+[D^-,\partial^+A^i]-[D^j,F^{ji}]=0\;.
\end{eqnarray}
We assume that when the parton traverses the medium, it induces a perturbation $\delta A^\mu$ of the strong medium field $A_0$ (linear response):
\begin{equation}
A^\mu=A^\mu_0+\delta A^\mu\;.
\end{equation}
Similarly for the conserved current
 \begin{equation}
J^\mu=J^\mu_0+\delta J^\mu\;.
\end{equation}
Keeping only terms in the Yang-Mills equations which are linear in the fluctuation  $\delta A^\mu$:
\begin{eqnarray}
&&
-\partial^+(\partial_\mu \delta A^\mu)=\delta J^+,
\nonumber\\
&&
\square \delta A^i-2ig[A_0^-,\partial^+\delta A^i]=\partial^i(\partial_\mu \delta A^\mu),
\nonumber\\
&&
\square\delta A^--2ig[A_0^-,\partial^+\delta A^-]=\delta J^-+2ig[\partial^iA_0^-,\delta A^i]+\partial^-(\partial_\mu \delta A^\mu)\nonumber\\
&&\quad\quad\quad\quad\quad\quad\quad \quad\quad\quad \quad\quad -ig[A^-_0,\partial_\mu \delta A^\mu].
\label{eq:YM}
\end{eqnarray}
The parton current obeys the conservation relation
\begin{equation}
\partial^+\delta J^-+D^-\delta J^+=\partial^+\delta J^-+\partial^-\delta J^+-ig[A_0^-,\delta J^+]=0.\label{eq:Dj}
\end{equation} 
With the initial condition $\delta J^+(x^+=t_0)=\delta(x^-)\rho(\x_\perp)$, where $t_0$ is the source production time, namely the production time of the hard parton in nucleus-nucleus collisions. In writing this current, we assume that the hard projectile propagates in the $x^+$ direction of the light cone; therefore, its interaction with the medium is eikonal: it only gets a color precession when passing through the medium. The solution of (\ref{eq:Dj}) reads
\begin{eqnarray}
\delta J^+&=&U(x^+,t_0,\x_\perp) \delta(x^-)\rho(\x_\perp)\theta(x^+-t_0)\;,\nonumber\\
\delta J^-&=&-\theta(x^+)\rho(\x_\perp)\delta(x^+-t_0)\;,
\end{eqnarray} 
where $U$ is a Wilson line in the adjoint representation of the
gauge group:
\begin{equation}
U(x^+,t_0,\x_\perp)\equiv {\cal T}_+\, \exp \left[ig\int_{t_0}^{x^+}
dz^+\; A_0^{-a}(z^+,\x_\perp)T_a\right]\; ,
\end{equation}
where ${\cal T}_+$ denotes the time ordering of the integrals along $z^+$. The current $\delta J^-$ corresponds to the propagation of an antiparticle moving in the opposite direction of the hard parton we are interested in. It is necessary to get a conserved current.
Making use of the the first equation of (\ref{eq:YM}) (seen as a constraint because it contains no time derivative) in the last equation for $\delta A^-$ we get :
\begin{equation}
\square\delta A^--2ig[A_0^-,\partial^+\delta A^-]=2ig[\partial^iA_0^-,\delta A^i]+\delta J^--\frac{1}{\partial^+}\left(\partial^-\delta J^+-ig[A^-_0,\delta J^+]\right),\label{A-}
\end{equation}
Thanks to the current conservation (\ref{eq:Dj}), $\delta J^+$ cancels out in right hand side of equation (\ref{A-}). Finally, we simplify equations (\ref{eq:YM}) which reduce to 
\begin{eqnarray}
&&
\square \delta A^i-2ig[A_0^-,\partial^+\delta A^i]=-\frac{\partial^i}{\partial^+}\delta J^+,
\nonumber\\
&&
\square\delta A^--2ig[A_0^-,\partial^+\delta A^-]=2ig[\partial^iA_0^-,\delta A^i]+2\delta J^-.
\label{eq:YM2}
\end{eqnarray}
The first equation of (\ref{eq:YM}) is solved leading to \footnote{In light cone gauge, we can forget about the $\delta A^-$ since it plays no role in the gluon spectrum.}: 
\begin{equation}
\delta A^i(x)=-\int d^4z\theta(z^-)G(x,z)\partial^i(U(z^+,t_0,\z_\perp)\rho(\z_\perp))\theta(z^+-t_0)\; .
\label{eq:field}
\end{equation}
The gluon propagator $G$ is the retarded Green's function obeying the equation of motion:
\begin{equation}
(\square_x -2ig(A_0^-.T)\partial_x^+)G(x,y)=\delta(x-y)\;,\label{eq:Green}
\end{equation}
with the initial condition, the free retarded propagator
\begin{equation}
G^0(x,y)=\frac{1}{2\pi}\theta(x^+-y^+)\theta(x^--y^-)\delta((x-y)^2) =\int \frac{d^4p}{(2\pi)^4}\frac{e^{-ip.(x-y)}}{p^2+i\epsilon p^+}.
\end{equation}
 The propagation of the emitted gluon, contrary to the hard projectile, is not necessarily eikonal.
Eq. (\ref{eq:field}) has a simple diagrammatic representation, shown in fig. (\ref{fig1}). The color precession of the source before the gluon emission is accounted for by the $U$ that multiplies $\rho$; while the rescatterings of the gluon after it is emitted are hidden in the Green's function $G$. The component $\delta A^-$ can be extracted from the constraint (first equation in (\ref{eq:YM})), but it is not relevant for gluon production since only transverse polarizations are physical. The integral over $z^-$ is restricted to the positive values: this comes from the term $(1/\partial^+)\delta J^+$ (which contains a $\delta(z^-)$) appearing in the second equation in (\ref{eq:YM}) after the substitution of $\delta A^-$ from the constraint.\\
The radiated gluon transverse field (\ref{eq:field}) is the main result of this section, it contains, as we will show, the physics of proton-nucleus collisions at high energy and of the induced radiative gluon spectrum in nucleus-nucleus collisions. 
\section{Induced radiative spectrum in A-A collisions} 

\begin{figure}[hbtp]
\begin{tabular}{c  c }

\resizebox*{!}{3cm}{\includegraphics{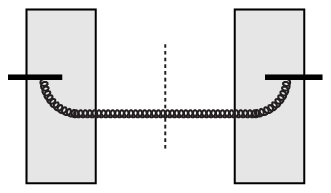}} &\resizebox*{!}{3cm}{\includegraphics{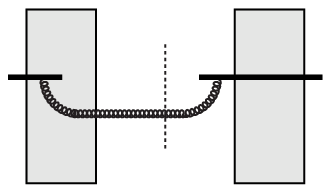}} \\
      \qquad (a)& \qquad(b)
\end{tabular} \caption{The diagrammatic representation of the two first terms in (\ref{eq:dI}). 
} \label{fig3}
\end{figure}

We assume that the fast projectile is a hard parton produced in a nucleus-nucleus collision at $t_0$, and take the origin of times when the parton enters the medium. In coordinate space, (for $x^+>L$) the gluon field (\ref{eq:field}), amputated of its final free propagator, reads    
\begin{eqnarray}
 \square_x \delta A^i(x)&=&-\theta(x^-)\partial^i_x(U(x^+,t_0;\x_\perp)\rho(\x_\perp))\nonumber\\
 &-&\int d^4z\theta(z^-)\theta(L-z^+)\theta(z^+-t_0)\delta(x^+-L)\nonumber\\
&\times&2\partial^+_xG(x,z)\partial^i_z(U(z^+,t_0;\z_\perp)\rho(\z_\perp)),\;\label{eq:ampfield}
\end{eqnarray}
where we used in the second term the following remarkable property of the Green's function (valid for $y^+<L<x^+$)\cite{BGV1}: 
\begin{equation}
G(x,y)=\int\limits_{z^+=L} dz^-d^2\z_\perp\,
G(x,z)\,2\partial_z^+ G(z,y)\; .
\label{eq:GreenId}
\end{equation}
In equation (\ref{eq:ampfield}), the first term corresponds to gluon emission occurring after the hard parton left the medium and the second term corresponds to gluon emission before the hard parton left the medium (if the parton is produced out side the medium, this emission could also occur before it enters the medium).
The invariance by translation, with respect to the - variables, of this Green's function is due to the fact that the medium field $A^-_0$ is independent of $x^-$, this can be seen in eq. (\ref{eq:Green}). This invariance implies that $G$ depends on the coordinates only via $x^--y^-$.\\
 It is useful to introduce a new Green's function, defined as 
\begin{equation}
{\cal G}_\omega(x^+,\x_\perp;y^+,\y_\perp)=2\int dl^- \partial_x^+G(x^+,\x_\perp;y^+,\y_\perp;l^-=x^--y^-)e^{il^-\omega}. 
\end{equation}
It is easy to verify that ${\cal G}_\omega$ is a Green's function of the two-dimensional Schr\"odinger operator:
\begin{equation}
(i\partial^ -+\frac{{\bs\partial}_\perp^2}{2\omega}+ig(A_0^-.T)){\cal G}_\omega(x^+,\x_\perp;y^+,\y_\perp)=i\delta(x^+-y^+)\delta(\x_\perp-\y_\perp)\;,
\end{equation}
therefore, it can be written in terms of a path integral for a quantum particle of mass $\omega$ moving in a potential:
\begin{eqnarray}
{\cal
G}_\omega(x^+,\x_\perp;y^+,\y_\perp)&=&\int{\cal
D}\r_\perp(\xi)\exp\left[\frac{i\omega}{2}\int^{x^+}_{y^+}d\xi
{\dot\r}_\perp^2(\xi)\right]\nonumber\\
&&\quad\quad\quad\times U(x^+,y^+;\r_\perp),
\end{eqnarray}
where $\r_\perp(x^+)=\x_\perp$ and $\r_\perp(y^+)=\y_\perp$. This path integral describes the Brownian motion of the emitted gluon in the transverse plane. Therefore, the emitted gluon follows a non-eikonal trajectory inside the medium \footnote{Since the medium is static these are the only non-eikonal corrections to gluon propagation.}. We recover the eikonal case by taking $\omega\rightarrow \infty$ or equivalently by assuming $x^+-y^+\rightarrow 0$: 
\begin{equation}
{\cal
G}_\omega(x^+,\x_\perp;y^+,\y_\perp)=\delta(\x_\perp-\y_\perp)\theta(x^+-y^+)U(x^+,y^+;\x_\perp),
\end{equation}
The Fourier transform of (\ref{eq:ampfield}) gives
\begin{eqnarray}
&&q^2\delta A^i(q)=\frac{ie^{iq^-L}}{\omega+i\varepsilon}\int d^2 \x_\perp e^{-i\q_\perp.\x_\perp}\big[\frac{q^i}{q^-+i\varepsilon}U(L,t_0;\x_\perp)\rho(\x_\perp)\nonumber\\
&&+ \int d^2\z_\perp \int_{t_0}^L dz^+ {\cal G}_\omega (L,\x_\perp;z^+,\z_\perp) \partial^i_z(U(z^+,t_0;\z_\perp)\rho(\z_\perp))\big ]\;.\label{eq:FTA}
\end{eqnarray}
The amplitude for the gluon radiation reads
\begin{equation}
{\cal M}_{(\lambda)}=q^2\delta A^i(q)\epsilon_{(\lambda)}^i(q)\;,\label{eq:amp}
\end{equation} 
where the gluon is taken on-shell  i.e.  $q^2=0$ (or $q^-=\q_\perp^2/2\omega$) . 
To get the average number of the produced gluons, we square the
amplitude and sum over the polarization vectors with the help of
the completeness identity
\begin{equation}
\sum_{\lambda}\epsilon^i_{(\lambda)}(\q)\epsilon^{j\ast}_{(\lambda)}(\q)=-g^{ij}.
\end{equation}
In the same way as in the CGC treatment we have to perform the average over the sources. First, for the fast parton, we write 
\begin{equation}
\left< \rho^a(\x_\perp)\rho^b(\x'_\perp)\right>_p=\mu^2_p(\x_\perp)\delta^{ab}\delta(\x_\perp-\x'_\perp)\;,
\end{equation}
$\mu^2_p(\x_\perp)$ is the parton charge density in the transverse plane. \\
From (\ref{eq:amp}) (or equivalently (\ref{eq:dN})), the gluon spectrum reads  
\begin{eqnarray}
&&\omega\frac{d\langle N \rangle}{d\omega d^2\q_\perp}= \frac{1}{16\pi^3} \sum_\lambda|{\cal M}_\lambda|^2=
\frac{1}{(2\pi)^3\omega}\Re e\int d^2\x_\perp\int d^2\y_\perp  e^{-i(\x_\perp-\y_\perp).\q_\perp}\nonumber\\
&&\times\int_{t_0}^L dz^+
\Big[\frac{1}{\omega}\int d^2\z_\perp\mu^2_p(\z_\perp)\int_{t_0}^{z^+}dz'^+\nonumber\\
&&\times
\left<{\bf tr}U(z^+,z'^+,\z_\perp)\partial^i_{z'}{\cal
G}^{\dag}_\omega(L,\y_\perp;z'^+,\z'_\perp)\partial^i_z{\cal
G}_\omega(L,\x_\perp;z^+,\z_\perp)\right>\vert_{\z_\perp=\z'_\perp}\nonumber\\ &&-2\mu^2_p(\y_\perp)\frac{q^i}{\q_\perp^2}\left<{\bf tr }U^\dag(L,z^+,\y_\perp)\partial^i_y{\cal
G}_\omega(L,\x_\perp;z^+,\y_\perp)\right> \Big]\nonumber\\
&&+\frac{2(N_c^2-1)}{(2\pi)^3\q_\perp^2}\int d^2 \x_\perp \mu^2_p(\x_\perp).\nonumber\\
\label{eq:dI}
\end{eqnarray}
For a single parton produced at $\x_\perp={\bf 0}_\perp$ it reduces to  
\begin{equation}
\mu^2_p(\x_\perp)=\frac{g^2C_R}{N_c^2-1}\delta(\x_\perp),
  \end{equation}
where $R=A$ for a gluon and $R=F$ for a quark.\\
In eq. (\ref{eq:dI}), $\x_\perp$ end $\y_\perp$ are the coordinates of the emitted gluon in the transverse plan.
The first term in (\ref{eq:dI}) corresponds to the probability of producing the gluon inside the medium, whereas the second term corresponds to the interference between the amplitudes for producing the gluon outside (after the parton left the medium),-first term in (\ref{eq:FTA})-, and inside the medium,-second term in (\ref{eq:FTA})-; this is illustrated in fig. \ref{fig3}. In the second term, note that the transverse coordinates of the emitted gluon and of the hard parton are the same (see the first term in (\ref{eq:FTA}) where the final gluon free propagator has been amputated). The last term is the probability of radiating a gluon in the vacuum, it has to be removed to get the medium induced gluon spectrum. This formula is quite general, in the sense that we do not specify the nature of the medium, so one still has to find a model for averaging over the medium sources. \\
Now we will show that this formula leads to the well known BDMPS-Z-W spectrum, in the case of uncorrelated sources. This approximation assumes that the scattering centers at different times in the medium are independent\footnote{This approximation holds when the range of one scattering center is much smaller than the mean free path of the radiated gluon and of the hard parton inside the medium.}, therefore the medium sources can be treated as Gaussian:
\begin{equation}
\left\langle \rho_0^a(x^+,\x_\perp)\rho_0^b(y^+,\y_\perp)\right\rangle=n(x^+)\delta(x^+-y^+)\delta^{ab}\delta(\x_\perp-\y_\perp),\label{eq:rho-gauss}
\end{equation}
Namely, that amounts to choose the statistical weight ${\cal W}$ as follows
\begin{equation}
{\cal W}[\rho_0]=\int {\cal D}[\rho_0]\exp\left\lbrace -\frac{1}{2}\int dx^+d^2\x_\perp\frac{\rho^a_0(x^+,\x_\perp)\rho^a_0(x^+,\x_\perp)}{n(x^+)}\right\rbrace ,
\end{equation}
where $n(x^+)$ is the medium scattering center density at the time $x^+$.
The corresponding correlator for $A^-_0$, given by (\ref{eq:A-}), reads:   
\begin{equation}
\left\langle A_0^{-a}(x^+,\x_\perp)A_0^{-b}(y^+,\y_\perp)\right\rangle=n(x^+)\delta(x^+-y^+)\delta^{ab}\gamma(\x_\perp,\y_\perp),\label{eq:AVgauss}
\end{equation} 
where $\gamma(\x_\perp,\y_\perp)$ can be expressed as follows:
\begin{equation}
\gamma(\x_\perp,\y_\perp)=\frac{1}{(2\pi)^2}\int \frac{d^2 \k_\perp}{\k_\perp^4}e^{i(\x_\perp-\y_\perp).\k_\perp},
\label{eq:gamma}
\end{equation} 
and is related to the dipole cross-section by the relation\footnote{for more details on Wilson line averages in the gaussian approximation see ref.\cite{BGV2,KovnW}}  
\begin{equation}
\sigma(\x_\perp-\y_\perp)=\frac{C_A}{2}\left[\gamma(\x_\perp,\x_\perp)+\gamma(\y_\perp,\y_\perp)+2\gamma(\x_\perp,\y_\perp)\right].
\end{equation}
In this approximation the average of two Wilson lines is a color singlet:
\begin{equation}
\left<U_{ac}(z^+,z'^+,\x_\perp)U_{cb}^\dag(z^+,z'^+,\y_\perp)\right>=\frac{\delta_{ab}}{N_c^2-1}\left<{\bf tr}U(z^+,z'^+,\x_\perp)U^\dag(z^+,z'^+,\y_\perp)\right>,
\end{equation}
and the two-point function reads
\begin {equation}
\left<{\bf tr } U(z^+,z'^+,\x_\perp)U^\dag(z^+,z'^+,\y_\perp)\right>=\exp\left[-\frac{1}{2}\int^{z^+}_{z'^+}d\xi n(\xi) \sigma(\x_\perp-\y_\perp)\right]\label{eq:UU}
\end {equation}
The second term in (\ref{eq:dI}) is easily evaluated leading to (see Appendix A for details):
\begin{eqnarray}
&&\frac{1}{N_c^2-1}\left<{\bf tr }{\cal
G}_\omega(z'^+,\y_\perp;z^+,\x_\perp)U^\dag(z'^+,z^+,\z_\perp)\right>\nonumber\\
&&\quad\quad\quad\quad={\cal K}_\omega(z'^+,\y_\perp-\z_\perp;z^+,\x_\perp-\z_\perp)\nonumber\\
&&\quad\quad\quad\quad=\int{\cal
D}\r_\perp(\xi)\exp\left[\int^{z^+}_{z'^+}d\xi (\frac{i\omega}{2}
{\dot\r}_\perp^2(\xi)-\frac{1}{2}n(\xi) \sigma(\r_\perp))\right],
\end{eqnarray}
where $\r_\perp(z'^+)=\y_\perp-\z_\perp$ and $\r_\perp(z^+)=\x_\perp-\z_\perp$.\\
Using the following property of the Green's functions, equivalent to (\ref{eq:GreenId}),
\begin{equation}
{\cal G}_\omega(x^+,\x_\perp;y^+,\y_\perp)=\int\limits_{y^+<u^+<x^+} d^2 \u_\perp {\cal G}_\omega(x^+,\x_\perp u^+,\u_\perp){\cal G}_\omega(z^+,\u_\perp;y^+,\y_\perp),
\end{equation}
and the locality in time of the source average, the first term in (\ref{eq:dI}) can be factorized as follows 
\begin{eqnarray}
&&\frac{1}{N_c^2-1}\left<{\bf tr}U(z^+,z'^+,\z_\perp){\cal
G}^{\dag}_\omega(L,\y_\perp;z'^+,\z'_\perp){\cal
G}_\omega(L,\x_\perp;z^+,\z_\perp)\right>\nonumber\\
&&\quad\quad\quad\quad=\frac{1}{N_c^2-1}\int d^2 \u_\perp \left<U^{ab}(z^+,z'^+,\z_\perp){\cal
G}^{\dag bc}_\omega(z^+,\u_\perp;z'^+,\z'_\perp) \right>\nonumber\\
&&\quad\quad\quad\quad\quad\quad\times\left<{ \cal
G}^{\dag cd}_\omega (L,\y_\perp;z^+,\u_\perp){ \cal
G}^{da}_\omega(L,\x_\perp;z^+,\z_\perp)\right>,\nonumber\\
&&\quad\quad\quad\quad=\frac{1}{(N_c^2-1)^2}\int d^2 \u_\perp \left<{\bf tr}U(z^+,z'^+,\z_\perp){\cal
G}^{\dag}_\omega(z^+,\u_\perp;z'^+,\z'_\perp) \right>\nonumber\\
&&\quad\quad\quad\quad\quad\quad\times\left<{\bf tr}{ \cal
G}^{\dag}_\omega(L,\y_\perp;z^+,\u_\perp){ \cal
G}_\omega(L,\x_\perp;z^+,\z_\perp)\right>.
\end{eqnarray}
Putting everything together, and following the calculations in Appendix A, we recover the well-known induced radiative gluon spectrum (see for instance \cite{BDMPS6}, where it is shown to be equivalent to Wiedemann's formulation \cite{Wied4})
\begin{eqnarray}
&&\omega\frac{ d\langle N\rangle}{d\omega d^2 \q_\perp}=\frac{\alpha_sC_R}{(2\pi)^2\omega}2\Re e\int_{t_0}^L dz^+\int d^2 \u_\perp e^{-i\q_\perp.\u_\perp}\nonumber\\
&&\times\Big[ \frac{1}{\omega}\int_{t_0}^{z+} dz'^+e^{-\frac{1}{2}\int_{z'^+}^L  d\xi n(\xi)\sigma(\u_\perp)} {\bs\partial}_{\perp y}. {\bs\partial}_{\perp u}{\cal K}_\omega(z^+,\u_\perp;z'^+,\y_\perp=0) \nonumber\\
&&-2\frac{\q_\perp}{\q_\perp^2}. {\bs\partial}_{\perp y}{\cal K}_\omega(L,\u_\perp;z^+,\y_\perp=0)\Big].
\end{eqnarray}

\section{gluon production in proton-nucleus collisions}
In this section, we will re-derive the gluon production in proton-nucleus collisions in the high energy limit \cite{GM,BGV1,KovchM3,DumitM1,KovchT1}. We end up by a discussion on the validity of the eikonal approximation at RHIC. 
\subsection{Gluon production in the high energy limit}
\begin{figure}[hbtp]
\begin{tabular}{c  c }

\resizebox*{!}{4cm}{\includegraphics{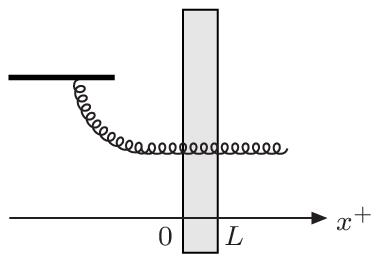}} &\resizebox*{!}{4cm}{\includegraphics{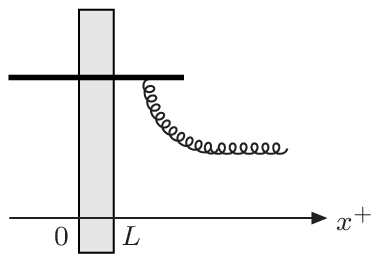}} \\
      \qquad (a)& \qquad(b)
\end{tabular} \caption{The two diagrams contributing to gluon production in proton-nucleus collisions in the limit $L\rightarrow 0$. In this limit the gluon is eikonal when passing through the nucleus, and the gluon emission inside the nucleus is neglected. 
} \label{fig2}
\end{figure}

At high energy the nucleus is Lorentz contracted, so that we take the limit $L\rightarrow 0$, and put $t_0=-\infty$. As a consequence the produced gluon is eikonal during the interaction time $L$, and the retarded gluon propagator (\ref{eq:Green}) simply reads
\begin{equation}
G(x,y)=\frac{1}{2}\theta(x^+-y^+)\theta(x^--y^-)\delta(\x_\perp-\y_\perp)U(x^+,y^+,\x_\perp).\;
\end{equation}
Thus, the gluon field (\ref{eq:field}), when amputated of its final free propatagor, reduces to 
\begin{eqnarray}
\square \delta A^i(x)&=&
2\delta(x^+)\delta(x^-)(U-1)\frac{\partial^i}{{\bs\partial}^2_\perp}
\rho(\x_\perp)
\nonumber\\
&-&
\theta(x^-)\theta(-x^+) \partial^i \rho(\x_\perp)
-\theta(x^-)\theta(x^+) \partial^i(U\rho(\x_\perp))\; ,\label{eq:fieldpA}
\end{eqnarray}
where $U\equiv U(+\infty,-\infty;\x_\perp)$. Then the Fourier transform gives
\begin{eqnarray}
-q^2 \delta A^i(q)&=&
-q^2 A_{\rm proton}^i(q)
\nonumber\\
&+&
2i\int \frac{d^2\k_{1\perp}}{(2\pi)^2}
\left[\frac{q^i}{2(q^++i\varepsilon)(q^-+i\varepsilon)}
-\frac{k_1^i}{k_{1\perp}^2}
\right]
\nonumber\\
&& \qquad\qquad\times \rho(\k_{1\perp})
\left[U(\k_{2\perp})-(2\pi)^2\delta(\k_{2\perp})\right]\;,\label{eq:fieldpA}
\end{eqnarray}
where $\k_{2\perp}\equiv \q_\perp-\k_{1\perp}$ and
$A_{\rm proton}^i(q)$ is the Fourier transform of the gauge field
of a proton alone, i.e. the Fourier transform of
eq.~(\ref{eq:fieldpA}) taking $U=1$. The two terms in (\ref{eq:fieldpA}) are illustrated in fig. (\ref{fig2}). This expression
leads to the standard result for gluon production in
proton-nucleus collisions.
\subsection{Validity of the eikonal approximation }

The center of mass energy per nucleon at RHIC is  200 GeV. It corresponds to a Lorentz contraction factor of about $\gamma=\sqrt{s}/2m_p\simeq 100$. For a nuclear radius $R_A\simeq 6.5$ fm, we end up with the estimate $L\simeq 0.5$ GeV$^{-1}$.
For the eikonal approximation to be valid, the gluon production time has to be much bigger than the maximum interaction time:
\begin{equation}
t_{prod}\sim\frac{\omega}{\q_\perp^2}\gg L . 
\end{equation}
At mid-rapidity, at RHIC, $ \omega\sim q_\perp\sim Q_s\sim$ 1GeV. We get $t_{prod}$ of order of $L$.  Therefore, one probably has  to go beyond the eikonal approximation for the gluon propagator to describe mi-rapidity data at RHIC. \\
The corrections to the eikonal approximation are taken into account in formula (\ref{eq:dI}) for gluon radiation. Even if (\ref{eq:dI}) has been derived in the framework of final state interactions in nucleus-nucleus, it is applicable for gluon production in proton-nucleus collisions (assuming the medium to be a nucleus and the projectile to be a proton, and taking the projectile production time $t_0=-\infty$). This is a possible phenomenological application for mid-rapidity hadron production at RHIC. \\
At forward rapidity, $\omega\sim q_\perp e^\eta$, therefore, $\omega$ is enhanced by a large factor (exponential of the rapidity) leading to a much larger gluon production time compatible with the eikonal approximation.

\section{The relation between $\hat q$ and $Q_s$}
The transport coefficient $\hat q$ which characterizes the density of the scattering centers (gluons in the medium) in the medium produced in nucleus-nucleus collisions is defined to logarithmic accuracy by $\frac{1}{2}{\hat q}(\xi)\r_\perp^2=n(\xi)\sigma(\r_\perp)$ \cite{KovnW2}. Using eq.  (\ref{eq:UU}), we can write the following relation:
\begin{equation}
\frac{1}{N_c^2-1}\left <{\bf tr}U^\dagger(x^+,y^+;\z_\perp)U(x^+,y^+;\z'_\perp)\right>=\exp\left[-\frac{1}{4}\int^{x^+}_{y^+}d\xi {\hat q}(\xi)(\z_\perp-\z'_\perp)^2\right].
\end{equation}
Whereas ${\hat q}$ is local in time, the saturation scale $Q_s$ is a global quantity defined as \cite{KovnW}: 
\begin{equation}
\frac{1}{N_c^2-1}\left <{\bf tr}U^\dagger(L,0;\z_\perp)U(L,0;\z'_\perp)\right>=\exp\left[-\frac{1}{4}Q_s^2(\z_\perp-\z'_\perp)^2\right],
\end{equation}
where $L\equiv 2R_A$ is the nuclear diameter.
From this formal analogy it becomes clear that the saturation scale appears as an initial condition for the transport coefficient in nucleus-nucleus collisions as suggested in \cite{Baier}
\begin{equation}
{\hat q}(\xi=0)\sim {\hat q}_{cold}\sim Q_s^2/2R_A.
\end{equation}
Obviously the transport coefficient increases due to the strong interactions of the freed gluons, from cold to hot matter. This mechanism is not discussed in this work, however, the medium dynamics could be encoded in the time evolution of the medium field, which could lead eventually to thermalization.
 \section{Summary}
In this paper, we calculate the gluon spectrum of a high energy and point-like projectile passing through a dense medium, presenting a compact and simple derivation based on the solution of the Yang-Mills equations in the light cone gauge of the fast projectile. We reproduce in a straightforward way the gluon spectrum in proton-nucleus collisions at high energy and the induced gluon spectrum in the final state of nucleus-nucleus collisions. We also discuss the validity of the eikonal approximation in gluon production in proton-nucleus collisions at RHIC, and argue that one could have non-negligible non-eikonal contributions at mid-rapidity gluon production. As an application of our formula for gluon production beyond the eikonal approximation (\ref{eq:dI}) to phenomenology, it could be interesting to see whether this complete solution leads to an appreciably different answer for hadron production spectra and for the Cronin peak at RHIC. Finally, we recall the relation between the saturation scale in initial state interactions and the transport coefficient in final state interactions, as the transition from cold to hot matter in nucleus-nucleus collisions.  \\
 
{\Large{\bf Acknowledgements}}\\

We would like to thank R. Baier, D. Dietrich, A. Dumitru, F. Gelis and D. Schiff for helpful discussions and for careful reading of the manuscript.      

\appendix 

\section{Averages over medium sources}
\setcounter{equation}{0}
\renewcommand{\theequation}{A-\arabic{equation}}
In this appendix we shall derive analytic expressions for some useful averages over the medium sources in the Gaussian approximation. 
\begin{eqnarray}
&&{\bf A}=\frac{1}{N_c^2-1}\left< {\bf tr}{\cal G}_\omega(x^+,\x_\perp;y^+,\y_\perp)U^\dag(x^+,y^+;\z_\perp)\right>\nonumber\\
&&\quad\quad\quad\quad=\int{\cal
D}\r_\perp(\xi)\exp\left[\frac{i\omega}{2}\int^{x^+}_{y^+}d\xi
{\dot\r}_\perp^2(\xi)\right]\nonumber\\
&&\quad\quad\quad\quad\times\frac{1}{N_c^2-1}\left < {\bf tr}U(x^+,y^+;\r_\perp(\xi))U^\dagger(x^+,y^+;\z_\perp)\right>,
\end{eqnarray}
where $\r_\perp(x^+)=\x_\perp$ and $\r_\perp(y^+)=\y_\perp$.
In the gaussian approximation the two-point function can be easily evaluated and gives
\begin{eqnarray}
&&\frac{1}{N_c^2-1}\left < {\bf tr}U(x^+,y^+;\r_\perp(\xi))U^\dagger(x^+,y^+;\z_\perp)\right>=\nonumber\\
&&\quad\quad\quad\quad\exp\left[-\frac{1}{2}\int^{x^+}_{y^+}d\xi n(\xi) \sigma(\r_\perp-\z_\perp)\right].
\end{eqnarray}
Now we define 
\begin{equation}
{\cal K}_\omega (x^+,\x_\perp;y^+,\y_\perp)=
 \int{\cal
D}\r_\perp(\xi)\exp\left[\int^{x^+}_{y^+}d\xi (\frac{i\omega}{2}
{\dot\r}_\perp^2(\xi)-\frac{1}{2}n(\xi) \sigma(\r_\perp))\right] .
\end{equation}
Making $\v_\perp=\r_\perp-\z_\perp$, we get 
\begin{equation}
\frac{1}{N_c^2-1}\left< {\bf tr} {\cal G}_\omega(x^+,\x_\perp;y^+,\y_\perp)U^\dag(x^+,y^+;\z_\perp)\right>={\cal K}_\omega (x^+,\x_\perp-\z_\perp;y^+,\y_\perp-\z_\perp).
\end{equation}
Now let evaluate the following useful average 
\begin{eqnarray}
&&{\bf B}=\frac{1}{N_c^2-1}\left< {\bf tr} {\cal G}^\dag_\omega(x^+,\x_\perp;y^+,\y_\perp){\cal G}_\omega(x^+,\z_\perp;y^+,\z'_\perp)\right>\nonumber\\
&&=\int{\cal
D}\r_\perp(\xi)\int{\cal
D}\u_\perp(\xi)\exp\left[\frac{i\omega}{2}\int^{x^+}_{y^+}d\xi
({\dot\u}_\perp^2(\xi)-{\dot\r}_\perp^2(\xi))\right]\nonumber\\
&&\times\frac{1}{N_c^2-1}\left< {\bf tr}U^\dag(x^+,y^+;\r_\perp)U(x^+,y^+;\u_\perp)\right>.\nonumber\\
\end{eqnarray}
with the following boundary conditions: $\r_\perp(x^+)=\x_\perp$, $\r_\perp(y^+)=\y_\perp$, $\u_\perp(x^+)=\z_\perp$ and $\u_\perp(y^+)=\z'_\perp$.
Recalling that the two-point function depends only on $\r_\perp-\u_\perp$ we perform the following change of variables 
\begin{eqnarray*}
&&{\bs\alpha}_\perp=\u_\perp-\r_\perp,\\
&&{\bs\beta}_\perp=\u_\perp+\r_\perp,
\end{eqnarray*}
yielding
\begin{equation}
{\bf B}=\int{\cal
D}{\bs\alpha}_\perp(\xi)\int{\cal
D}{\bs\beta}_\perp(\xi)\exp\left[\int^{x^+}_{y^+}d\xi
(\frac{i\omega}{2}{\dot{\bs\alpha}}_\perp(\xi){\dot{\bs\beta}}_\perp(\xi)-\frac{1}{2}n(\xi)\sigma({\bs\alpha}_\perp))\right],
\end{equation}
Now we can perform the $\bs\beta$ integration yielding to a $\delta$-function constraining the $\bs\alpha$ variable to a straight line as following 
\begin{equation}
\l_\perp(\xi)=\frac{1}{(x-y)^+}\left[(\xi-y^+){\bs\alpha}_\perp(x^+)+(x^+-\xi){\bs\alpha}_\perp(y^+)\right].
\end{equation}
It is straightforward to show, for instance by discretizing the path integral,  that
\begin{eqnarray}
{\bf B}&=&\left(\frac{\omega}{2\pi (x-y)^+}\right)^2\nonumber\\
&\times&\exp\left[
\frac{i\omega}{2(x-y)^+}((\x-\y)_\perp^2-(\z-\z')_\perp^2)-\frac{1}{2}\int^{x^+}_{y^+}d\xi n(\xi)\sigma(\l_\perp(\xi))\right].\nonumber\\
\end{eqnarray}
We shall calculate the Fourier transform appearing in (\ref{eq:dI}) 
\begin{eqnarray}
{\bf C}=\int d^2\x_\perp \int d^2\z_\perp {\bf B} \quad e^{-i\k_\perp.(\x-\z)_\perp}.
\end{eqnarray}
Let 
\begin{eqnarray}
&&\v_\perp=(\x-\y-\z+\z')_\perp,\nonumber\\
&&\w_\perp=(\x-\y+\z-\z')_\perp,
\end{eqnarray}
We get 
\begin{eqnarray}
&& {\bf C}=\left(\frac{\omega}{2\pi (x-y)^+}\right)^2 e^{-i\k_\perp.(\y-\z')_\perp}\int d^2\v_\perp \int d^2\w_\perp e^{-i\k_\perp.\v_\perp} \nonumber\\
&&\quad\quad\quad\times\exp\left[
\frac{i\omega}{2(x-y)^+}\v_\perp.\w_\perp-\frac{1}{2}\int^{x^+}_{y^+}d\xi n(\xi)\sigma(\l_\perp(\xi))\right] ,
\end{eqnarray}
where now 
\begin{equation}
\l_\perp(\xi)=\frac{(\xi-y^+)}{(x-y)^+}\v_\perp+(\y-\z')_\perp.
\end{equation}
$\sigma(\l_\perp)$ is independent of $\w_\perp$, thus the integral over $\w_\perp$ yields a $\delta (\v_\perp)$ and we finally get 
\begin{equation}
{\bf C}=\exp\left[-i\k_\perp.(\y-\z')_\perp
-\frac{1}{2}\int^{x^+}_{y^+}d\xi n(\xi)\sigma((\y-\z')_\perp)\right].
\end{equation}
\bibliographystyle{unsrt}

\end{document}